\documentclass[aps,prd, preprintnumbers]{article}

\usepackage{graphicx, url, multicol, longtable,amsmath, amssymb}

\makeatletter

\newenvironment{makefigure}
{\def\@captype{figure}\vskip 12pt}
{}
\makeatother

\begin{document}

\title{Simulation of Cosmogenic Neutrino Spectra with the
  GZKFast Event Generator}
\author{\normalsize John A. Cairns$^1$\\
{\small \em $^1$ Department of Physics, The Ohio State University, Columbus,
OH 43210, USA}\\
{\small\tt cairnsj@mps.ohio-state.edu, john@2ad.com}}

\date{\normalsize May 25, 2006\\
\small (updated May 15, 2007)}
\maketitle\thispagestyle{empty}

\begin{abstract}
GZKFast is a low-cost astrophysical event generator designed to
simulate photohadron processes resulting from ultra high energy
cosmic ray fluxes.   GZKFast is an easy to use event generator which
specifically addresses issues relevant to cosmogenic neutrinos and the
ultra high energy (UHE) neutrino spectrum.   GZKFast injects UHE
particles into a simulated cosmic microwave background (CMB) 
using a Monte Carlo approach.  The interaction of each particle is 
simulated and the resulting detectable events are spooled to log
files.  Although GZKFast uses a simplified particle physics model
and limited accelerator data it generates event distributions which
are comparable with more thorough simulations such as
the event generator SOPHIA \cite{Muckeetal}.
\end{abstract}


\begin{multicols}{2}

\section{Advertisement}

GZKFast is a capable event generator which produces results
comparable with applications that incorporate far more substantial
particle physics kinematics and accelerator data.  
\vskip 6pt
GZKFast provides...
\begin{itemize}
\item A full featured simulation with rich run time configuration.
\item Monte Carlo event generation.
\item Management of individual astrophysical sources.
\item Individual event simulation and tracking including relevant
  particle kinematics.
\item Modular C++ design.
\item Histogram and extensive tabular output.
\item Practical standalone approach designed for ease of use,
  portability, and redistribution.
\item Reusable program library for easy incorporation into new projects.
\item GNU Library General Public License.
\end{itemize}

\section{Introduction}

This year the ``GZK Effect'' of Greisen \cite{Greisen} and Zatsepin
and Kuz'min \cite{Zatesepin} is celebrating its fortieth anniversary.
These authors independently realized that a uniform cosmic microwave
background would provide an optically thick background for the highest
energy cosmic rays.  Cosmic ray protons of sufficiently large energy would
be likely to interact with CMB photons over suitably large distances. 

Accelerator experiments predict that the important channels for
such interactions would be direct photopion production and the resonance
channels for the $\Delta$ \cite{Mucke}.

\begin{align} 
\label{pprxn}
p + \gamma \rightarrow n + \pi^+ \notag\\
p + \gamma \rightarrow \Delta^+ \rightarrow n + \pi^+\\
\ldots \rightarrow p + \pi^0\notag
\end{align}

The stable daughters of these processes may be observable in Earth
based detectors.  Therefore we wish to predict the expected fluxes of
these particles.

Four momentum invariance of these reactions implies that in the proton
rest frame \cite{Stecker}:

$$ s = M_p^2 + 2 M_p E_\gamma $$

where $M_p$ is the proton mass and $E_\gamma$ is the energy of the
photon in the lab frame.  Therefore the measured cross section for
photo-hadron interactions, $\sigma (s)$, determines the
likelihood for proton attenuation.  The expected attenuation length
can be characterized by a mean free path, $\lambda$.

$$ \lambda = {1\over n\sigma}$$

The mean free path also depends on the average density of CMB photons
$n_{avg}$.  The average density may be found by assuming the CMB is
characterized by a black body spectrum and integrating the Bose
distribution.  Including redshift, the average density of the CMB is given by (\ref{navg}).

\begin{equation}
\label{navg}
 n_{avg} =  (1+z)^3 {2\zeta(3)k_b^3 T^3\over\hbar^3 c^3 \pi^2}
\end{equation}

Here $\zeta(x)$ is the zeta function defined in statistical mechanics
texts and $T$ is the temperature of the microwave background today.
Using accepted values, at present this quantity is:

$$ n_{avg} = 410.5\, cm^{-3} $$

The cross section for (\ref{pprxn}) can be determined from the
Breit-Wigner formula using the measured data for the expected resonances.
The relativistic form can be found in
most introductory particle physics texts.

\begin{equation}
\label{bw}
\sigma(s) = \sigma_{max} {{M_0}^2\Gamma^2\over{(s-{M_0}^2)^2 +
    \Gamma^2{M_0}^2}}
\end{equation}

The peak of the resonance is scaled by $\sigma_{max}$ which may be
measured experimentally.  $M_0$ is the mass of the resonance, and
$\Gamma$ is the width. 

The mean attenuation length for a cosmic ray proton of a given
energy is the energy fraction of scattering times the attenuation length.

$$ L_0 = \left( {E\over \Delta E} \right) \lambda $$

For photopion production the energy fraction can be deduced from
kinematics, and averages roughly ${\Delta E/ E} = 15 \%$ in the
relevant energy regime.  We can estimate the mean attenuation length
of a $ 10^{20} \, eV $ cosmic ray proton, taking an average value of
$\sigma$, $500 \mu b$: 

$$ L_0 = \left( {E\over \Delta E} \right) (n \sigma)^{-1} \approx
10^{25} \, cm \approx 10 Mpc$$

A cosmic ray traveling a distance, $L$, would have a survival
probability less than $P_{survival}$. 

$$ P_{survival} (L) = 1.0 - \exp{(-L/L_0)} $$

Therefore, the flux of ultra high energy cosmic rays should be
significantly attenuated on cosmological scales.  The product of these
photohadron interactions is a flux of electrons, photons, protons,
neutrons and neutrinos.  However, only the neutrinos travel without
attenuation on these cosmological distances.   These \em cosmogenic
neutrinos \rm could be detected by Earth based experiments. 

It is a topical goal of the astrophysics community to measure and
characterize the flux of ultra high energy neutrinos. 
Still, no contemporary experiment is capable of measuring the
cosmogenic neutrino flux, so suitable Monte Carlo event generators
must be available to support ongoing research.  Here we present the
C++ code named {\tt gzkfast} as one possible avenue for simulation of
this process.

\section{Program Operation}

GZKFast is principally comprised of two components, a reusable library
and a command line event generator.  The reusable program library, \tt
libgzkparticle, \rm is written in C++ and designed to be modular and
portable across platforms.  The simple invocation program, \tt
gzkfast, \rm gives command line access to the key functionality provided in the
GZK library.  

When a user invokes the \tt gzkfast \rm program, a universe is created
with a simplified particle physics model.   Cosmic ray point sources
are inserted at random locations and managed by an \tt
EventGenerator \rm thread.   The event generator thread iterates through the
point sources and injects subsequent ultra high energy particles into
an \tt EvolutionThread \rm with momentum oriented towards the Earth. 

Cosmic ray events are managed in particle queues of the evolution
thread modules using a ``round robin'' strategy.   As a result
execution time is divided relatively equally between available threads.

Each evolution thread maintains a list of relevant \tt Space \rm
objects, including the \tt CMB, \rm and extragalactic space, or \tt
BFieldSpace. \rm  The evolution threads also maintain a list of
suitable Earth based \tt Detector \rm objects.    

Particle evolution continues as particles propagate through each of
the known spaces and are given a chance to be detected by known
detectors.  Basic particle kinematics are used for propagation.  In
the extragalactic $\bf B \rm$ field, assumed to be static and uniform
over distance $dx$, the kinematics of particle propagation are
governed by a few simple relations.

$${d\bf p\rm\over dt} = {e\over c} \bf v\times B$$
$$ \bf p \rm = \bf p \rm + {\rm d\bf p\over\rm dt}\delta t $$
$$ \bf x \rm = \bf x \rm + \bf v\rm\delta t$$
$$ \rm t = t + \delta t$$

The time step, $\delta t$, is defined based on the user specified
distance step, $dx$.  Particle interactions with CMB photons are evaluated
with an accept-reject strategy.  Then after propagation, each particle
is allowed to decay with probability $P_{decay}$.

$$ P_{decay}(t) = 1.0 - \exp{(-t/\tau)} $$

Decay products are reinserted into the evolution thread particle
queue.  If no decay or particle production takes place during
propagation, each known detector is tested to see if it can see the
particle.  If the particle ``hits'' the detector the particle becomes
an ``event.''  

Each event is logged to a relevant output location.  Particle
evolution continues until the user specified number of events have
been detected.  Once the specified number of events have been
recorded, event generation is halted.  Program execution then terminates
once all particle queues are empty. 

\section{Basic Results and Spectra}

Each run of \tt gzkfast \rm is configurable to allow the user to
explore different aspects of cosmogenic neutrino phenomenon.  With
suitable configuration it is easy to produce a wide variety of
relevant data.

GZKFast uses integrated data provided by the Particle Data Group
\cite{RPP} to sample the proton-photon cross section versus
$\sqrt{s}$.   Accelerator data provided in the PDG dataset are
linearly interpolated from point to point to produce a complete
distribution as depicted in Figure \ref{deltares}.   

Using (\ref{bw}) to determine the cross section for the $\Delta_{1232}$ and
$\Delta_{1600}$.  These data are
tabulated in the file specified by the ``cmb'' basename, with
``\_sigmaplot.dat'' appended.

\makefigure {
\centering
\label{deltares}
\includegraphics{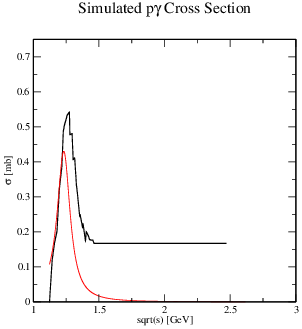}
\caption{The $p\gamma$ cross section measured in accelerator
  experiments (Black) \cite{RPP}.  The 1232
  MeV $\Delta$ resonance as determined from the Breit-Wigner formula (Red).
  The difference is taken to be the cross section for direct pion
  photoproduction.}
}

The cosmic microwave background is uniformly sampled as a black body
distribution at 2.725~K \cite{Bennett}.   These data are output in the
file specified by the ``cmb'' basename, with ``\_photonhist.dat''
appended.

\makefigure {
\centering
\includegraphics{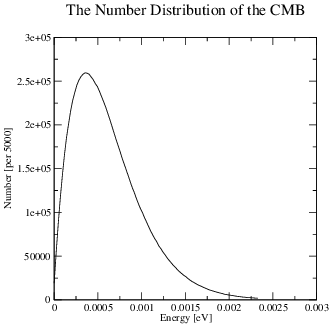}
\caption{The number of photons at a given energy for a 5000 $\nu$ event
  simulation.}
}

At program termination the \tt gzkfast \rm program produces a variety
of output spectra for the various processes that it tracked.  Most
importantly GZKFast is capable of producing neutrino spectra resulting from
super-GZK cosmic rays.   These data are output in the file specified
by the ``neutrino'' basename, with ``\_hist.dat'' or ``\_event.dat''
appended depending on if the data set is a histogram or event file.

\makefigure {
\centering
\includegraphics{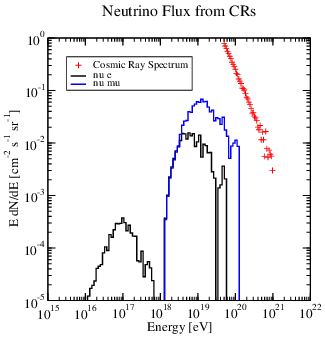}
\caption{The neutrino histogram gives the energy distribution of an expected
neutrino flux.}
}

GZKFast also provides individual event data for protons and neutrons,
neutrinos and photons.  These event data can be utilized in detector
simulation or since they also include right ascension and declination
information they can be used to produce sky maps with variable
sources, distances, and other configuration differences.



\end{multicols}

GZKFast source code is available to the public upon request.

\section{Acknowledgment}

I acknowledge John~Beacom, Matthew~D.~Kistler and
Michael~S.~Sutherland for discussions involving
this work.

\section{GZKFast Reference}
\subsection{Class Listing}
\begin{longtable}{lp{4.75in}}
\tt CMB \rm & The Cosmic Microwave Background Object samples the cross
section of $p\gamma$ and the distribution of a user specified black body
spectrum to determine if an interaction with a cosmic ray will
occur.\\
\tt CMBDist \rm & Sample the distribution of a black body radiation of a
  given temperature.\\
\tt Delta \rm & Program representation of a $\Delta$ (1232 MeV) particle.
Provides decay kinematics and particle properties.\\
\tt Delta1600 \rm & Program representation of a $\Delta$ (1600 MeV) particle.
Provides decay kinematics and particle properties.\\
\tt Detector \rm & Abstract virtual class interface for GZKFast
spherical detectors.   A detector is a volume of space which can be
\em hit \rm by a simulated particle.\\
\tt Electron \rm &  Program representation of an $e^-$ particle.
Provides decay kinematics and particle properties.\\
\tt EventGenerator \rm & Threaded object which iterates through cosmic
ray sources producing particle events from a predetermined energy
spectrum and inserting them into the particle queues.\\
\tt EvolutionThread \rm & Threaded object which iterates through a
specific particle event queue and propagates the particles through one
of an arbitrary number of spaces.  After the particle propagates, the
detectors are given a chance to detect it.\\

\tt G4Vector \rm & A 4 vector representation.\\
\tt GGuard \rm & Guard a critical section.\\
\tt GHistogram \rm & A linear scale histogram class. \\ 
\tt GLogHistogram \rm & A log scale histogram class. \\
\tt GMath \rm & Provides rudimentary numerical operations and random
number generation. \\
\tt GMatrix \rm & An object representation of an $n\times n$ matrix.\\
\tt GMutex \rm & An object to provide mutual exclusion across
platforms.\\
\tt GRunThread \rm & A C++ implementation of a Java-like ``java.lang.Thread''
object.\\
\tt GThread \rm & An object to provide rudimentary threading support
across platforms.\\
\tt GVector \rm & A vector of arbitrary size, based on \tt
std::valarray.\rm\\
\tt GVegas \rm & The Vegas Numerical Recipe.\\
\tt BFieldSpace \rm & Propagate a particle through a ``static'' magnetic
field using Hamiltonian formalism. \\
\tt Mu \rm & Program representation of an $\mu$ particle.
Provides decay kinematics and particle properties.\\
\tt Neutron \rm & Program representation of an neutron.
Provides decay kinematics and particle properties.\\
\tt Nu \rm & Program representation of a neutrino.
Provides decay kinematics and particle properties.\\
\tt NuDetector \rm & An instance of \tt Detector \rm which is capable of
detecting neutrinos.\\

\tt Particle \rm & An abstract virtual class interface for particle
objects. \\
\tt Photon \rm & Program representation of a photon.
Provides decay kinematics and particle properties.\\
\tt PhotonDetector \rm & An instance of \tt Detector \rm which is
capable of detecting photons.\\
\tt Pion \rm &  Program representation of a $\pi$.
Provides decay kinematics and particle properties.\\
\tt PiZero \rm &  Program representation of a $\pi^0$.
Provides decay kinematics and particle properties.\\
\tt Proton \rm &  Program representation of a $p^+$.
Provides decay kinematics and particle properties.\\
\tt ProtonDetector \rm & An instance of \tt Detector \rm which is
capable of detecting protons.\\

\tt ProtonSource \rm & A point source for cosmic ray protons.\\
\tt ProtonSpectrum \rm & An object to sample a power law energy
spectrum of cosmic ray protons. \\
\tt Source \rm & An abstract virtual class interface for cosmic ray
sources.\\
\tt Space \rm & An abstract virtual class interface for spaces.  A
space allows a particle to propagate.\\
\tt Sphere \rm & A geometric representation of a sphere.  Provides ray
intersection for detector operation.\\
\tt ThreeBodyDecay \rm & A generalized Monte Carlo of the reaction $A
\rightarrow B + C + D$.\\
\tt TwoBodyDecay \rm & A solution for two body decay in the ultra high
energy limit.   Decay products are collinear with parent particles.\\
\tt Universe \rm & Provides a frame for calculating redshift and
volume integrals.\\
\end{longtable}

\subsection{Command Line Reference}
\begin{longtable}{|l|l|l|p{2in}|}\hline
\bf Parameter & \bf Units & \bf Default & \bf Description \\\hline\rm
-sources & Number & 3 & The number of cosmic ray
    sources to add to the simulation. Sources are added randomly to a
    shell of size specified by -near and -far.\\\hline
-events & Number & 100 & The number of neutrino events to simulate before
stopping. \\\hline
-threads & Number & 2 & The number of asynchronous execution paths
simultaneously processing input events. \\\hline
-alpha & Number & -2.7 & Simulate a proton input spectrum of the form
$E^\alpha$. \\\hline
-low & EeV & 50 & The starting energy for the input proton
distribution.\\\hline
-hi & EeV & 50000 & The ending energy for the input proton
distribution.\\\hline
-near & Mpc & 150 & The distance to the nearest cosmic ray
  source.\\\hline 
-far & Mpc & 200 & The distance to the furthest cosmic ray
  source.\\\hline 
-dx & Kpc & 250 & The distance corresponding to one iteration or step
of the Monte Carlo integration.\\\hline
-rad & Kpc & 250 & The radius of the ``detector'' volume.   The
``detector'' is defined to be a volume of given radius centered about
the earth.   Any particle which will intersect the ``detector'' is
considered to be an ``event.'' \\\hline
-bfield & Gauss & 1e-9 & The maximum magnitude of the uniform
extragalactic magnetic field.\\\hline
-quality & Number & 5e-9 & The precision of the beta decay Monte
Carlo.\\\hline
-proton & File Name & ``proton'' & The base name of the files for writing
proton events and histogram output.\\\hline
-2nd & File Name & ``secondaryproton'' & The base name of the files for
writing secondary proton events and histogram output.\\\hline
-v & File Name & ``neutrino'' & The base name of the files for writing
neutrino events and histogram output.\\\hline
-cmb & File Name & ``cmb'' & The base name of the files for
  writing the cmb photon energy histogram and cross section sample
  data.\\\hline 
-photon & File Name & ``photon'' & The base name of the files for
writing photon events and histogram output.\\\hline
\end{longtable}

\subsection{Program Output}
\begin{verbatim}
$ gzkfast: -[ arguments ] ...
Simulate a flux of ultra high energy neutrinos from cosmic ray sources.

        -sources #    - The number of sources.
        -events #     - The number of events to simulate.
        -threads #    - The number of processor threads.
        -alpha #      - Simulate E^alpha spectrum.
        -low #        - Least energy [EeV].
        -hi #         - Highest energy [EeV].
        -near #       - Least source distance [Mpc].
        -far #        - Highest source distance [Mpc].
        -dx #         - distance step [Kpc].
        -rad #        - detector radius [Kpc].
        -bfield #     - B field strength [Gauss].
        -quality #    - The precision of Monte Carlo convergence.
        -proton file  - Name of file for input protons.
        -2nd    file  - Name of file for secondary protons.
        -v      file  - Name of file for neutrinos
        -cmb    file  - Name of file for cmb distributions
        -gamma  file  - Name of file for photons.
\end{verbatim}

\subsection{File Formats}

%

\begin{table}[h]
\begin{tabular}{cc}
\rm Energy [eV] &  \rm E dN/dE [$cm^{-2} s^{-1} sr^{-1}$] \rm\\
4.235324597282258e+16& 6.205560549060830e-05\\
4.829213438776459e+16& 5.812199626341997e-05\\
5.506379003919574e+16& 1.066115138632034e-04\\
\end{tabular}
\caption {Example neutrino\_hist.dat output.}
\end{table}

\begin{table}[h]
\begin{tabular}{cccccc}
\rm RA [deg] &  \rm Dec [deg] & \rm E [eV] & \rm
 $p_x$ [eV] & \rm  $p_y$ [eV] & \rm $p_z$ [eV]\rm\\
73.8507& -31.1954& 4.2491+16&2.6999+16& -7.5372+15&-3.1932+16\\ 
73.8507& -31.1954& 8.9698+18&5.6996+18& -1.5911+18&-6.7409+18\\ 
82.8030& -90.6093& 1.0938+16&-3.6303+15& 3.6691+15&-9.6441+15\\ 
82.8030& -90.6093& 1.4671+19&-4.8691+18& 4.9212+18&-1.2934+19\\ 
\end{tabular}
\caption {Example neutrino\_event.dat output.}
\end{table}


\end{document}